\documentclass
[
amssymb,prl,aps,twocolumn,floatfix,tightenlines,superscriptaddress]{revtex4}
\pdfoutput=1
\usepackage[arrow, matrix, curve]{xy}
\usepackage{graphics}
\usepackage{graphicx,color}
\usepackage{subfigure}
\usepackage{float}

\usepackage[ansinew]{inputenc}
\usepackage{graphicx}
\usepackage{amsmath}
\usepackage{amssymb}
\usepackage{dsfont} 
\usepackage{hyphenat}
\usepackage[percent]{overpic}

\begin{document}
\title{Coherent Time Evolution and Boundary Conditions of Two-Photon Quantum Walks}
\author{J. D. A. Meinecke}
\author{K. Poulios}
\affiliation{Centre for Quantum Photonics, H. H. Wills Physics Laboratory
and Department of Electrical and Electronic Engineering, University
of Bristol, Merchant Venturers Building, Woodland Road,
Bristol BS8 1UB, UK.}
\author{A. Politi}
\altaffiliation{Now at: Center for Spintronics and Quantum Computation, University of California, Santa Barbara, California 93106, USA}
\author{J. C. F. Matthews}
\author{A. Peruzzo}
\affiliation{Centre for Quantum Photonics, H. H. Wills Physics Laboratory
and Department of Electrical and Electronic Engineering, University
of Bristol, Merchant Venturers Building, Woodland Road,
Bristol BS8 1UB, UK.}
\author{N. Ismail}
\author{K. Wörhoff}
\affiliation{Integrated Optical Microsystems
Group, MESA+ Institute for Nanotechnology, University
of Twente, Enschede, Netherlands.}
\author{J. L. O'Brien}
\author{M. G. Thompson}
\email{Mark.Thompson@bristol.ac.uk}
\affiliation{Centre for Quantum Photonics, H. H. Wills Physics Laboratory
and Department of Electrical and Electronic Engineering, University
of Bristol, Merchant Venturers Building, Woodland Road,
Bristol BS8 1UB, UK.}

\date{\today}

\begin{abstract}
Multi-photon quantum walks in integrated optics are an attractive controlled quantum system, that can mimic less readily accessible quantum systems and exhibit behaviour that cannot in general be accurately replicated by classical light without an exponential overhead in resources. The ability to observe time evolution of such systems is important for characterising multi-particle quantum dynamics---notably this includes the effects of boundary conditions for walks in spaces of finite size. Here we demonstrate the coherent evolution of quantum walks of two indistinguishable photons using planar arrays of 21 evanescently coupled waveguides fabricated in silicon oxynitride technology. We compare three time evolutions, that follow closely a model assuming unitary evolution, corresponding to three different lengths of the array---in each case we observe quantum interference features that violate classical predictions. The longest array includes reflecting boundary conditions.
\end{abstract}
\maketitle

Random walks describe stochastic motion of a particle around a discrete space and
are widely applied as a statistical tool in areas ranging from genetics to economics. The quantum mechanical version---quantum walks---is an interference phenomenon of quantum particles that exhibits distinctly different dynamics. Quantum walks on many graphs exhibit ballistic propagation of the walker's probability distribution, while many trials of a classical random walk will build a distribution localised around the initial position \cite{ke-cotphys-44-307}---a 1-dimensional quantum walk is a typical example. For certain tasks, quantum walk dynamics are beneficial, and have inspired new quantum algorithms \cite{ch-pra-70-022314} and approaches to universal quantum computing by realising quantum walks on a finite set of graphs \cite{ch-prl-102-180501}; in this setting, quantum walks are traditionally thought of as being simulated on a quantum computer.
Quantum walks are also a physical phenomena in their own right, describing for example evolution in solid state systems \cite{bo-prl-91-207901,ch-prl-92-187902} and transport phenomena in biomolecules \cite{pl-njp-10-113019}---all such systems are finite and therefore exhibit either periodic, absorbing, reflecting or cyclic boundary conditions which have been treated theoretically, for example \cite{mue-prle-71-036128, be-arXiv-0609128}. 
 
There are two main models of quantum walks \cite{ke-cotphys-44-307}---continuous time and discrete time, of which single walker examples have been demonstrated experimentally using trapped ions \cite{sc-prl-103-090504,za-prl-104-100503,ka-sci-325-174}; optical resonators \cite{bo-pra-61-013410};  NMR \cite{ry-pra-72-062317,du-pra-77-042316}; single photons in bulk \cite{do-josab-22-499,br-prl-104-153602} and fibre optics \cite{sc-prl-104-050502,sch-prl-106-180403}; and laser light in coupled waveguide arrays \cite{pe-prl-100-170506}. The observable dynamics of single walkers can be described by either single particle quantum mechanics or classical wave mechanics \cite{kn-pra-68-020301}, as demonstrated by optical quantum walks using laser light. When multiple indistinguishable particles undergo a quantum walk, correlations occur that cannot in general be explained or mimicked with classical physics, without either sacrificing visibility in observable features \cite{br-prl-102-253904}---quantified by an inequality relating correlated outcomes---or introducing a factorially increasing number of experiments \cite{ke-pra-83-013808} or resources. The latter approach has been recently adopted with optical fibres and laser light \cite{sc-sci-336-55}. Two particle quantum walks \cite{br-prl-102-253904} have been demonstrated with photons guided in integrated optics: Indistinguishable photons have lead to quantum correlations in a planar \cite{pe-sci-329-1500} and a circular \cite{ow-njp-13-075003} array of evanescently coupled waveguides; polarisation entangled quantum walks have recently been demonstrated and shown to emulate the quantum interference statistics of fermions, bosons and the intermediate regime \cite{ma-arxiv-1106.1166,sa-prl-108-010502}.

Here we measure three distinct time steps of a two-photon continuous-time quantum walk in three
arrays of 21 waveguides, with identically designed coupling and propagation parameters, integrated on a single chip. 
The length of each array $z$ is varied, to provide the mechanism for observing multi-photon dynamics in time ($z=c\cdot t$ for speed of light $c$ in the structure).
The longest time step exhibits boundary effects from photons being reflected at the outer waveguides. In all three cases, we observe quantum interference that violates an inequality that relating correlations between different waveguides and assumes classical light as the input \cite{br-prl-102-253904}.

\begin{figure}
\includegraphics[width=0.85\linewidth]{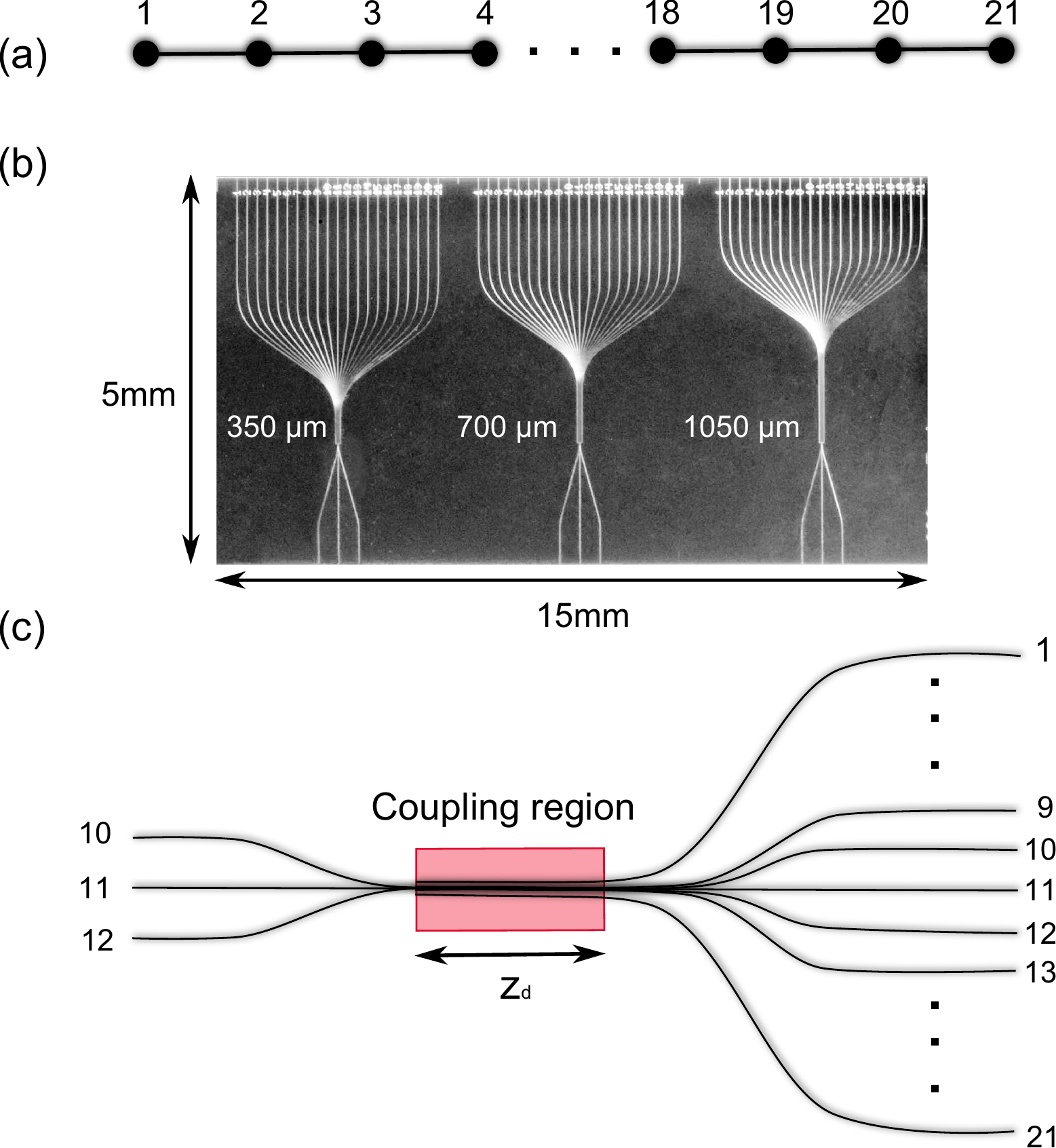}
\caption{Waveguide structure for measuring time evolution. (a) Graph representing the planar array of 21 waveguides. (b) Chip with the three measured waveguide arrays. One can see the three input waveguides for each array, followed by the narrow coupling region and the spreading to the output at the top of the chip. (b) Schematic of the waveguide structure; the waveguides run parallel within the coupling region and bend to the input and output separation uniformly.}
\label{chip}
\end{figure}

Photons propagating in a planar array of $N$ evanescently coupled single mode waveguides realise a quantum walk on a one dimensional lattice (see Fig. \ref{chip} (a) for N=21).
Assuming nearest-neighbour coupling from the evanescent fields between waveguides $j$ and $j\pm 1$ (amplitude $\kappa_{j,j\pm 1} = \kappa_{j\pm 1,j}$) and a waveguide propagation constant $\beta_j$, the Hamiltonian is modeled with
\begin{equation}
\hat{H}=c\sum_{j=1}^N{\beta_j a_j^{\dagger}a_j+\kappa_{j, j-1}a_{j-1}^{\dagger}a_{j}+\kappa_{j, j+1}a_{j+1}^{\dagger}a_{j}},
\label{hamiltonian}
\end{equation}
where $a_j^{\dagger}$ and $a_j$ are the bosonic creation and annihilation operators for a photon in waveguide $j$ and
for uniform arrays $\kappa_{i,j}$ and $\beta_j$ are constant.

Time evolution of a given state $\left|\psi(t)\right\rangle$ is determined by the unitary operator $U(t)=\exp(-iHt)$
according to $\left|\psi(t)\right\rangle=U(t)\left|\psi(t_0=0)\right\rangle$.
The transition amplitude from the initial state of a particle in waveguide $j$ ($a^\dagger_j\left|0\right\rangle =\left|1\right\rangle_j$) at time $t_0=0$ to the state of a particle in waveguides $j'$ ($a^\dagger_{j'}\left|0\right\rangle =\left|1\right\rangle_{j'}$) at time $t$ is then given by
$U_{j',j}(t)= \left\langle 1\right|_{j'}\exp(-iHt)\left|1\right\rangle_j$,
yielding the single particle transition probability
\begin{equation}
p_{j',j}(t)=\left|U_{j',j}(t)\right|^2,
\label{eqn:single}
\end{equation}
which also describes the photon density distribution one expects in experiments with coherent light.

Quantum correlations occur in the output state of the quantum walk of multiple indistinguishable particles \cite{br-prl-102-253904}.
From injecting a photon pair into waveguides $k$ and $j$, the probability to detect one photon in waveguide $j'$ coincident with detecting one photon in waveguide $k'$ is given by the two-photon correlation function \cite{ma-apb-60-s111}
\begin{equation}
\Gamma^{(j,k)}_{j',k'}(t)=\frac{1}{1+\delta_{j',k'}}\left|U_{j',j}(t)U_{k',k}(t)+U_{k',j}(t)U_{k,j'}(t)\right|^2.
\label{cor}
\end{equation}
Non-classical features emerge from interference of the complex probability amplitudes inside the $|\cdot|^2$. Expression (\ref{cor}) with unitary operator $U$ is the theoretical description to which we compare our experimental results.
The correlation function of distinguishable particles is treated using classical probability theory, multiplying and summing the possible single photon transition probabilities: $\Gamma'^{(j,k)}_{j',k'}(t)= p_{j',j}(t)p_{k',k}(t)+p_{j',k}(t)p_{k',j}(t)$.
To distinguish between two-photon quantum interference experiments and any classical treatment of light, reference \cite{br-prl-102-253904} defined the inequality
\begin{equation}
V_{j,k}(t)=\frac{2}{3}\sqrt{\Gamma'_{j,j}(t)\Gamma'_{k,k}(t)}-\Gamma'_{j,k}(t)<0
\label{v}
\end{equation}
which is only violated by the interference of indistinguishable photons \cite{pe-sci-329-1500}.

The quantum walk Hamiltonian (\ref{hamiltonian}) is implemented with waveguide arrays fabricated in silicon oxynitride ($\text{SiO}_{x}\text{N}_{y}$) with a designed refractive index contrast of $4.4\%$.
This high index contrast enables fabrication of micron sized single mode waveguides which are pitched at $2.8\mu\text{m}$ within the coupling region in order to achieve sufficient mode overlap for nearest neighbour coupling. The waveguides are designed with constant height of 0.6$\mu\text{m}$ and width of 1.8$\mu\text{m}$.
Three waveguide arrays with coupling region lengths $z=350\mu\text{m}$, $700\mu\text{m}$ and $1050\, \mu\text{m}$ were fabricated on a compact 5 mm long chip (Fig. \ref{chip}) and measured to simulate three values of time evolution parameter $t=z/c$.

Identical, horizontally polarised 808nm photons generated from type-I downconversion source (see Appendix) are launched into each of the three arrays
using a butt-coupled array of polarisation maintaining, single mode optical fibre with a standard pitch of $250\,\mu\textrm{m}$; this addresses the input of the array by bending the input waveguides to a separation of $250\, \mu\text{m}$ (with a maximum bend radius of 600$\mu\,\textrm{m}$). The output of the array is similarly addressed by bending the output waveguides to a separation of $125\,\mu\textrm{m}$ and butt-coupling the output chip to a second $250\,\mu\textrm{m}$ pitch array allowing correlated two-photon detection with commercially available fibre-coupled avalanche photo-diode single photon counting modules, across even and odd labeled waveguides. Two-photon correlations were measured by recording pairs of detection within a $5 \text{ns}$ window as a coincidence event.
Coincidence logic was performed with field programmable gate arrays. Photon-number resolving detection was performed non-deterministically using fibre beam splitters.

\begin{figure}
	\centering
		\includegraphics[width=0.95\linewidth]{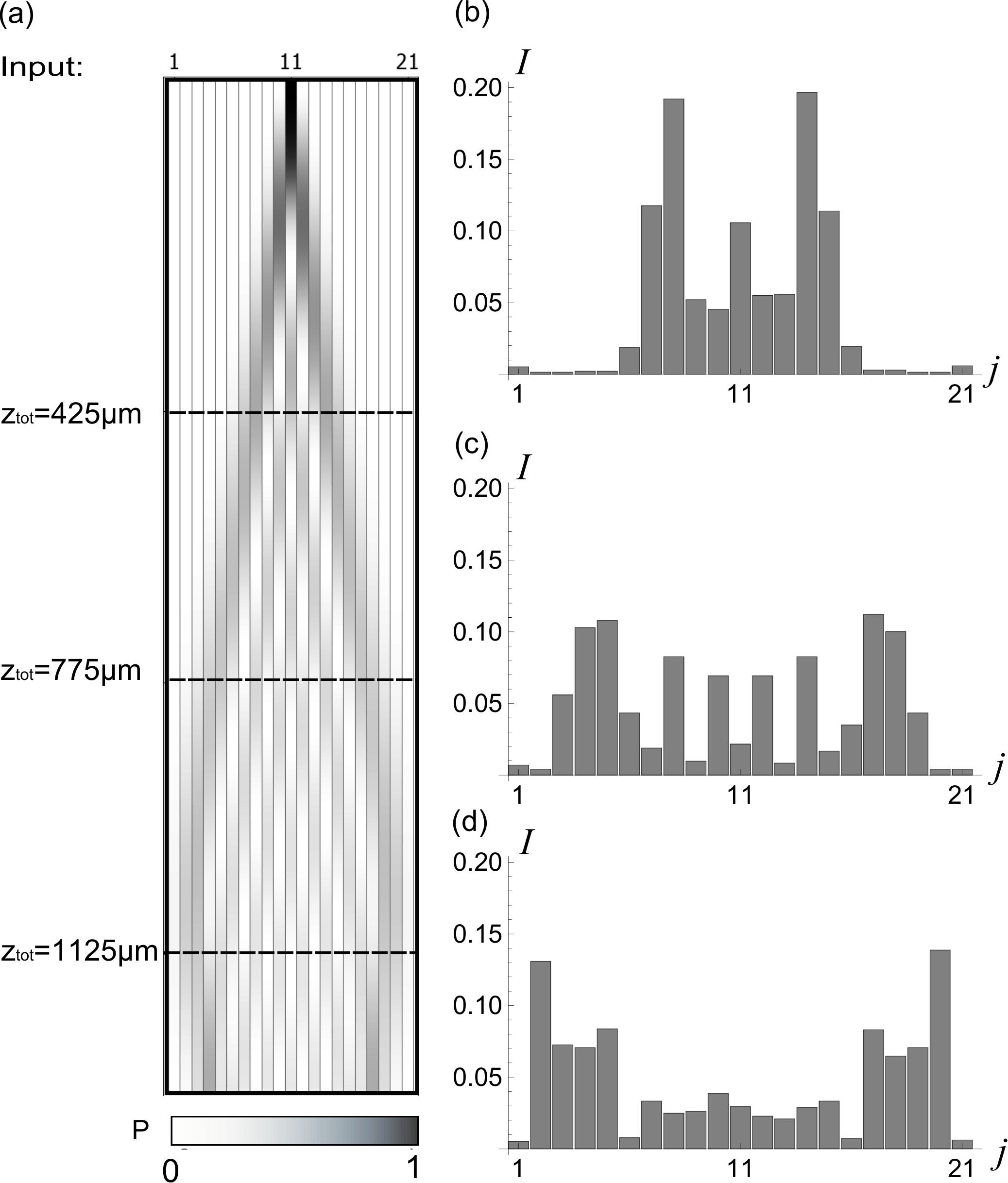}
  \caption{Single photon evolution. (a) Simulated probability distribution for a single photon walk starting in the middle of the array at waveguide number 11. (b-d) Three measured intensity distributions $I$ normalized to the total intensity for the arrays of (b) 350 $\mu\text{m}$ (c) 700$\mu\text{m}$ and (d) 1050 $\mu\text{m}$.}
\label{single}
\end{figure}

The single photon distribution $p_{j',j}(t)$ was characterised with 810nm laser light injected into the central waveguide (input waveguide 11). 
Fig. \ref{single} shows the single photon distribution measured with a power intensity meter, for the three different array lengths in comparison with the theoretical prediction calculated according to equation (\ref{eqn:single}).

Two photon correlations $\Gamma_{j',k'}(t)$ were measured from injecting indistinguishable photon pairs into the two pairs of input waveguides ($\{11,12\}$ and $\{10,12\}$) that provide the most overlap throughout the device, for each of the three simulated time evolutions.
Indistinguishability of the photons was verified by performing a Hong-Ou-Mandel type experiment \cite{ho-prl-59-2044} with an average visibility of $V = 98.4 \pm 0.5\%$.
The correlation matrices for inputs $\{10,12\}$ for the three propagation lengths are shown in Fig. \ref{cormats} (a)-(c) (correlations for inputs $\{11,12\}$ given in the Appendix).

Each of the 121 measured  two-photon coincidence possibilities (from all 231 possible combinations) were measured over 15 minute integration periods and are
corrected for relative detector efficiency.
We quantify the agreement between the theoretical calculations $\Gamma^{th}_{j',k'}(t)$ and the experimental data $\Gamma^{exp}_{j',k'}(t)$ with the
similarity 
$S=\left(\sum_{j',k'}{\sqrt{\Gamma^{exp}_{j',k'}\cdot \Gamma^{th}_{j',k'}}}\right)^2/\sum_{j',k'}{\Gamma^{exp}_{j',k'}}\sum_{j',k'}{\Gamma^{th}_{j',k'}}$.

In order to derive the values of the entries of the Hamiltonian (\ref{hamiltonian}) of our waveguide structure we use standard coupled mode theory \cite{lifante}
according to which the propagation constant of waveguide $j$ is given by $\beta_j=n\frac{2\pi}{\lambda}$ where $n$ is the effective refractive index and $\lambda$ the vacu\-um wavelength.
The coupling constant $\kappa$ between two identical waveguides can be deduced from the propagation constants of the two normal modes $\beta_e$ and $\beta_o$ of the coupled system as $\kappa=(\beta_o-\beta_e)/2$.
We use a waveguide mode solver (FIMMWAVE) to calculate the corresponding propagation constants.
The uniform array with the designed waveguide dimensions can be modeled with coupling and propagation constants of $\kappa=0.00489 \mu\text{m}^{-1}$ and $\beta=11.40462 \mu\text{m}^{-1}$. The spreading regions at the input and output add an effective coupling length of $75 \mu\text{m}$ to each array length, so that the total coupling lengths are given by $z_{tot}=t_{tot}\cdot c=425 \mu\text{m}, 775 \mu\text{m} \ \text{and} \ 1125 \mu\text{m}$.

Maximising the similarity of the measured two-photon distribution with theoretical predictions we obtain a Hamiltonian (Eq. (\ref{hamiltonian})) with
\begin{equation*}
\beta_{i}= \begin{cases}
11.385 \ \textrm{for} \ i=1,21 \\
11.403  \ \textrm{for} \ i=2,20\\
11.397  \ \textrm{else},
\end{cases}
\kappa_i= \begin{cases}
0.0049 \ \textrm{for} \ i=1,20 \\
0.0058  \ \textrm{for} \ i=2,19\\
0.00513  \ \textrm{else}.
\end{cases}
\end{equation*}
where $\kappa_i=\kappa_{i,i+ 1}=\kappa_{i+1,i}$.

This suggests fabrication deviation since boundary effects like an additional self coupling term added to the propagation constant of outermost waveguides are a magnitude smaller than the derived perturbations \cite{cha-jos-2-6}.
The resulting similarities between theory and experiment are summarised in Table \ref{sim}. 

\begin{table}
\begin{tabular}{|c|c|c|c|}\hline
$z_d$ &$z_{tot}$   &  S, input 11,12 & S, input 10,12\\\hline
350$\mu$m &425$\mu$m	&		$94.0 \pm0.2\%$ & 90.0$ \pm0.2\%$\\\hline
700$\mu$m &775$\mu$m &  90.3$ \pm0.1\%$  &90.1$ \pm0.1 \%$ \\\hline
1050$\mu$m &1125$\mu$m & 89.6$ \pm0.1\%$ & 90.7$ \pm0.1\%$ \\\hline
\end{tabular}
\caption{Similarities for different time steps and 2-photon input combinations.}
\label{sim}
\end{table}

\begin{figure*}
	\centering
		\includegraphics[width=.86\textwidth]{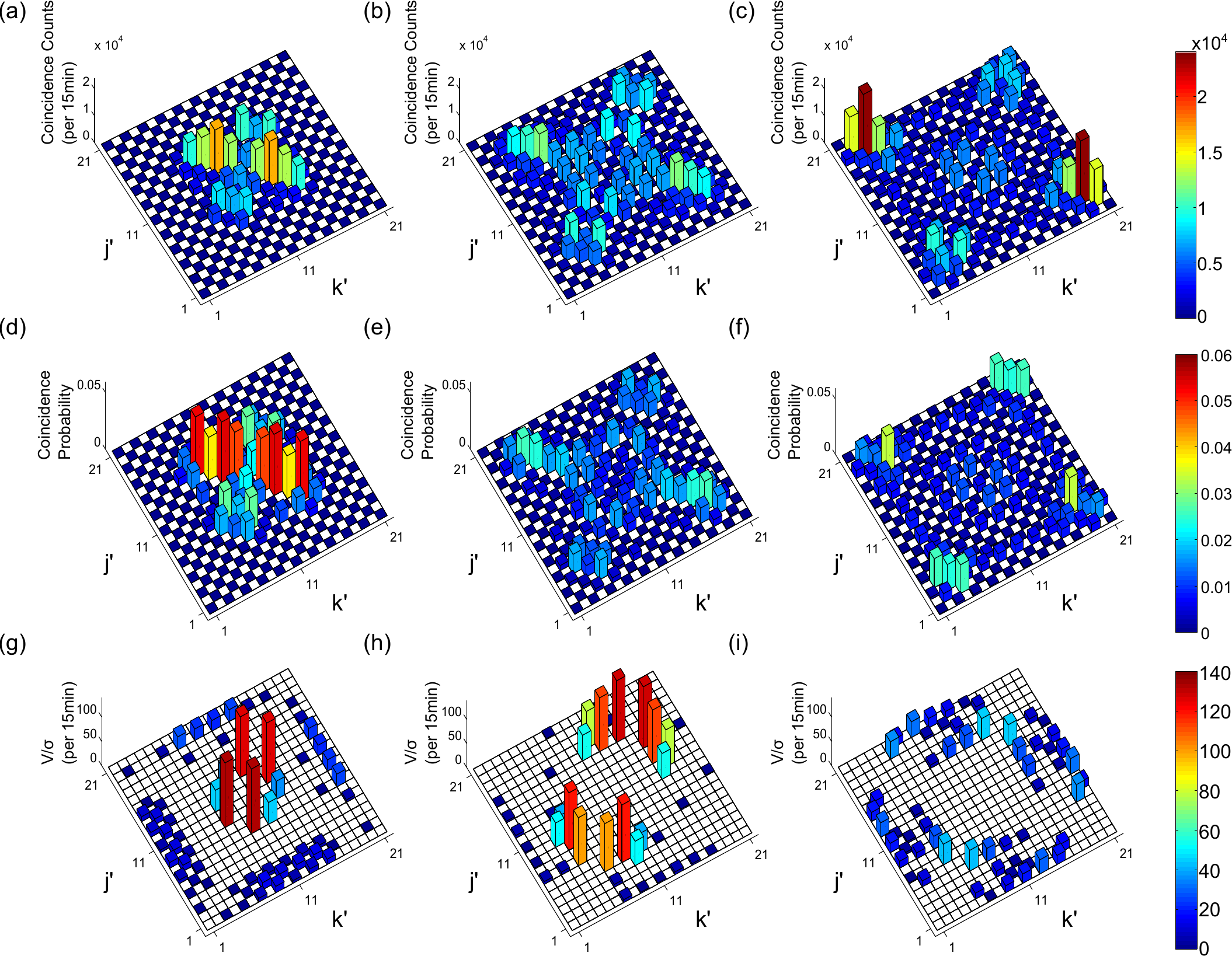}
\caption{Two-photon data. Experimental [(a)-(c)] and theoretical correlations [(d)-(f)] for detecting simultaneously one photon in waveguide $j'$ and one in waveguide $k'$ when injecting photon pairs in waveguide 10 and 12. (g)-(i) shows the violations of equation (\ref{v}) in units of standard deviations.}
    \label{cormats}
\end{figure*}

\begin{figure}
\includegraphics[width=0.85\linewidth]{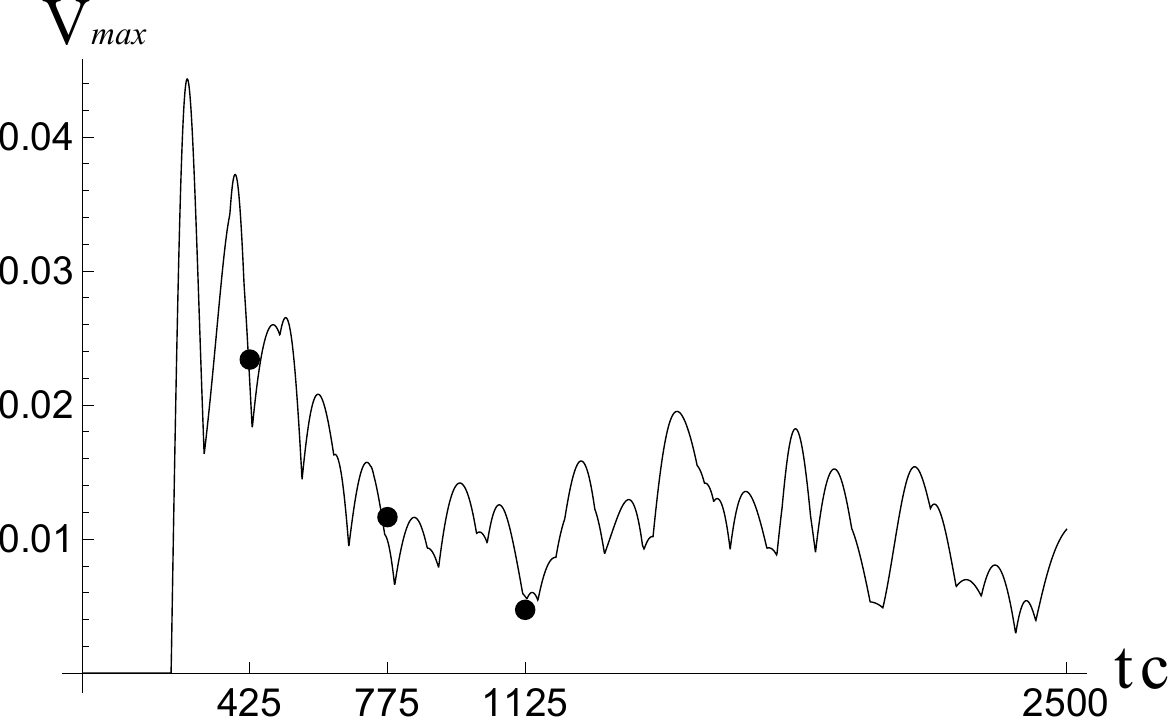}
\caption{Time evolution of maximum violation $V_{max}$ for input 10 and 12 in comparison with measured maximum violations (points). Errorbars on experimental data are smaller than data points. Theoretical and experimental correlation matrices used to calculate the violations are normalized as probability distributions.}
\label{vt1}
\end{figure}

We verify the quantum behaviour of the photons at all three time steps through the violation of the classical inequality (\ref{v}) as depicted in Fig. \ref{cormats}(g)-(i).
The array length of 700 $\mu$m shows the strongest violation with a maximal value of 128 standard deviations, while the longest array (1050 $\mu$m) shows comparatively small  violations of a maximum of 45 standard deviations.
This behaviour is in agreement with a theoretical study of the time dependence of the violation (Fig. \ref{vt1}). We calculate the violations for the theoretical correlation matrices (for all 121 measurable two-photon coincidences) and plot the maximum value.

Within the time interval relevant for our experiment the maximum violation decreases with time as the photons distribute over the waveguides. For longer times, as the photons reached the boundaries of the array and are back-reflected, the maximum violation shows an oscillatory pattern.
The local minima and maxima for the maximum violation reflect the changing phase relation between the matrix elements of the unitary time evolution operator $U_{j,k}$.

In summary, we have observed the time evolution of a two-photon continuous time quantum walk on an array of 21 evanescently coupled waveguides. The three measured time-steps agree with theoretical prediction made assuming coherent unitary evolution; this is evidence of coherent evolution of multi-photon states within integrated optical circuits. However, full characterisation of time evolution can be achieved via state tomography and as the size of quantum walks increase, schemes that approximately determine the quantum process using resources efficiently could be implemented \cite{cr-ncom-1-12}.
We have observed that the magnitude with which two-photon quantum interference violates classical prediction varies over the three measured time steps.
Our numerical simulations support these fluctuations, and specifically predict that it is possible to have zero violations, despite genuine two-photon interference.
In further array designs the pitch and width of the outer waveguides can be adapted to model various boundary gradients. 

\begin{small}
\noindent We thank J. Carolan and D. Fry for helpful discussion. This work was supported by EPSRC, ERC, IARPA, QUANTIP, PHORBITECH, and NSQI. J.C.F. M. is supported by a Leverhulme Trust Early Career Fellowship. A. Pe. holds a Royal Academy of Engineering Research Fellowship. J.L.OB. acknowledges a Royal Society Wolfson Merit Award.
\end{small}

\appendix*

\section{Appendix}

\begin{figure*}
	\centering
		\includegraphics[width=.95\textwidth]{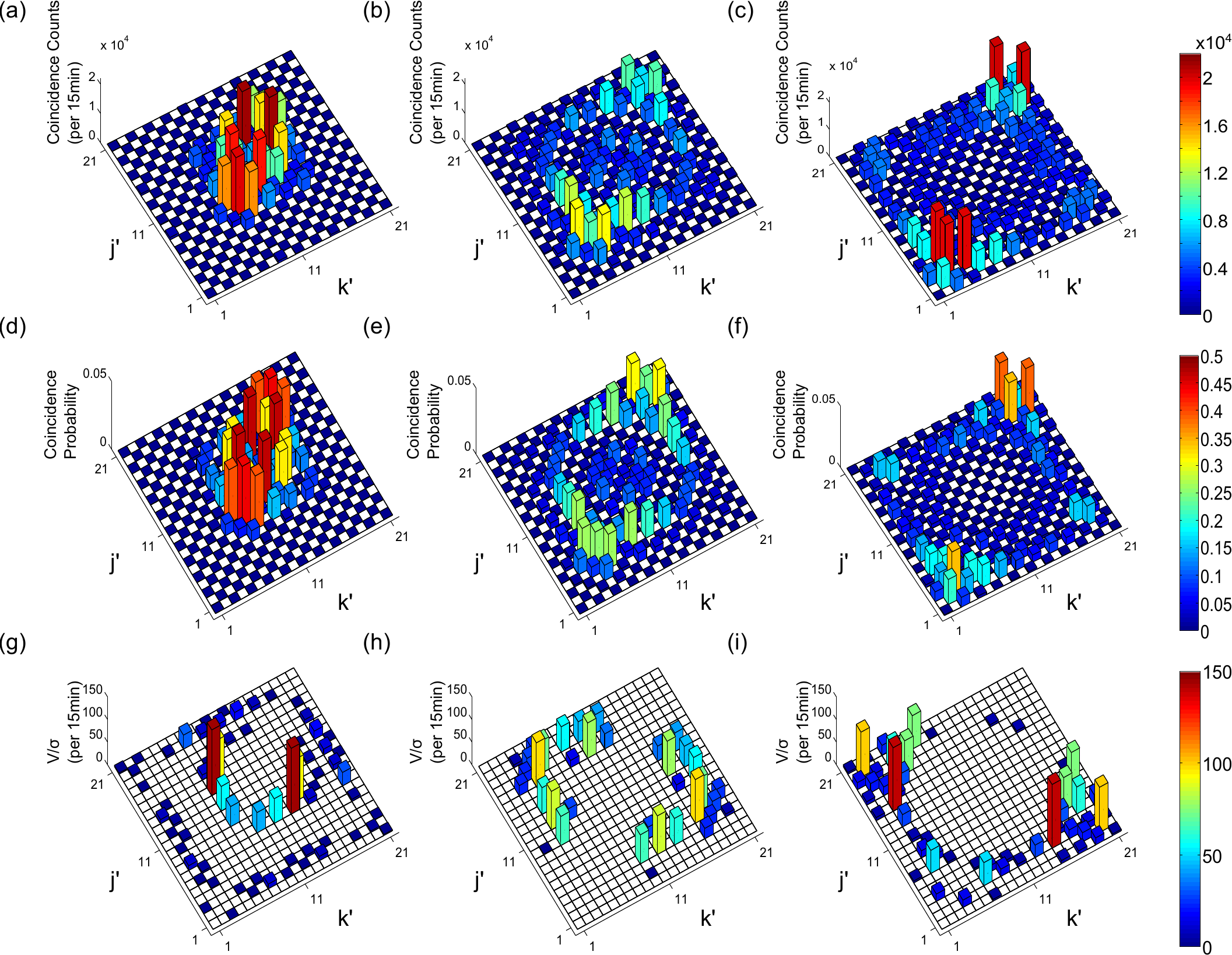}
\caption{Experimental [(a)-(c)] and theoretical correlations [(d)-(f)] for detecting simultaneously one photon in waveguide $j'$ and one in waveguide $k'$ when injecting photon pairs in waveguide 11 and 12. (g)-(i) shows the violations of equation (4) in the main text in units of standard deviations.}
\label{cormats011}
\end{figure*}

\begin{figure}
\includegraphics[width=0.75\linewidth]{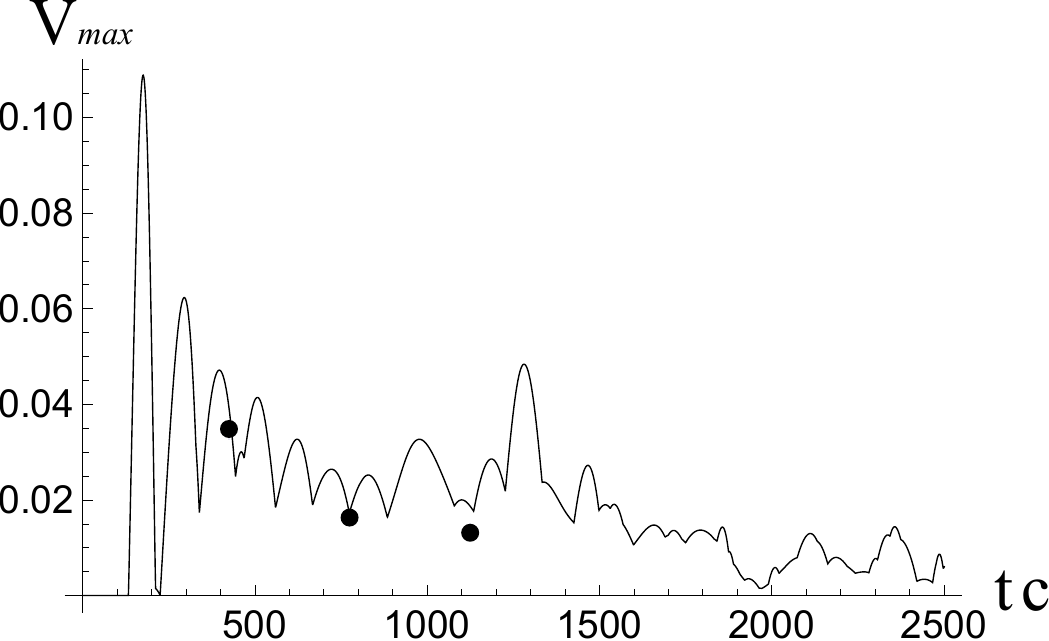}
	\caption{Time evolution of maximum violation $V_{max}$ for input 11 and 12 in comparison with measured maximum violations (points). Errorbars on experimental data are smaller than data points. Theoretical and experimental correlation matrices used to calculate the violations are normalized as probability distributions.}
\label{vt}
\end{figure}

\textbf{Experimental results for input waveguides 11 and 12.} Here we present the experimental results for a second input combination when injecting photon pairs in neighbouring waveguides numbered 11 and 12 (see Fig. 1 (c) in main text). Fig. \ref{cormats011} shows the correlation matrices and violations measured.
Quantum interference results in distinctly different correlations compared to the input combination of waveguides 10 and 12. In all three time evolutions we observe a generalized bunching behaviour, the two particles are likely to go to the same side of the array.
Fig. \ref{vt} compares theoretically predicted time evolution of the maximum violation with our measurements. 
Compared to the input $\{10,12\}$, the photons wavefunctions start overlapping at an earlier time, so violations can be observed at an earlier time (around  $t\cdot c=130\mu\text{m}$). Despite genuine two-photon interference, the maximum violation goes down to zero at $t\cdot c=230\mu\text{m}$. 

\textbf{Two-photon down conversion source.} The photon pairs were produced via type I spontaneous parametric down conversion in a 2-mm-thick $\chi^2$ nonlinear bismuth borate $\text{BiB}_3\text{O}_6$ crystal.
A continuous-wave diode laser ($404\text{nm}$) pumped the crystal with 60 mW to create degenerate photons at a wavelength of $\lambda=808 \text{nm}$.
Interference filters ($3.1\text{nm}$) ensured spectral indistinguishability before we launched the light into two polarization-maintaining fibres (PMF), which were butt-coupled to the chip with refractive index matching liquid.

\begin{figure}
	\centering
		\includegraphics[width=0.73\linewidth]{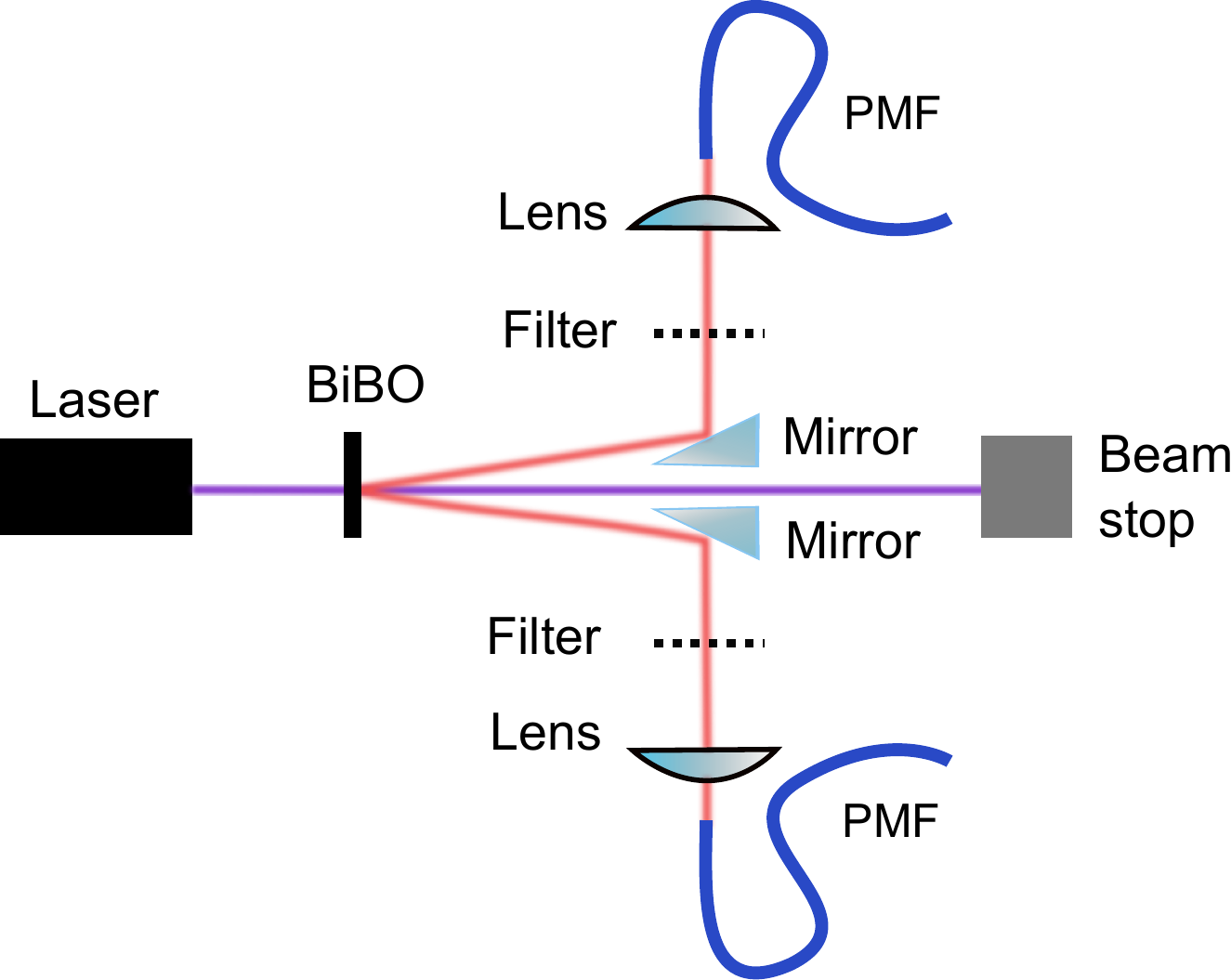}
	\caption{Schematic of the type-I down conversion source used to generate (unentangled) photon pairs.}
	\label{fig:setup}
\end{figure}

\end{document}